\input phyzzx \def\e{\adveq\eqno{\rm (\chapterlabel\the\equanumber)}}

\def\adveq{\global\advance\equanumber by 1}
\def\myeq{{\rm \chapterlabel\the\equanumber}}

\def\dim{\mathop{\rm dim}\nolimits}

\def\semidirect{\mathrel{\raise0.04cm\hbox{${\scriptscriptstyle |\!}$
\hskip-0.175cm}\times}}

\def\mod{\mathop{\rm mod}\nolimits}

\def\ref#1{$^{[#1]}$}

\def\r#1{$[\rm#1]$}

\def\e{\adveq\eqno{\rm (\chapterlabel\the\equanumber)}}

\def\adveq{\global\advance\equanumber by 1}
\def\myeq{{\rm \chapterlabel\the\equanumber}}

\def\dim{\mathop{\rm dim}\nolimits}

\def\semidirect{\mathrel{\raise0.04cm\hbox{${\scriptscriptstyle |\!}$
\hskip-0.175cm}\times}}

\def\mod{\mathop{\rm mod}\nolimits}

\def\ref#1{$^{[#1]}$}

\def\r#1{$[\rm#1]$}

\def\half{{1\over2}}

\overfullrule=0pt
\date{April,  2017}
\date{April, 2017}
\titlepage
\title{Generalized Rogers Ramanujan Identities for Twisted Affine Algebras }
\author{Arel Genish and Doron Gepner}
\vskip20pt
\line{\it\hfill  Department of Particle Physics, Weizmann Institute, Rehovot, Israel\hfill} 

\abstract
The characters of parafermionic conformal field theories  are given by the string functions
of affine algebras, which are either twisted or untwisted algebras. Expressions for these characters as generalized Rogers Ramanujan algebras
have been established  for the untwisted affine algebras. However,
we  study the  identities for the string functions of twisted affine Lie algebras. 
Conjectures for the string functions  was proposed by Hatayama et al., for the unit fields,
which expresses the string functions as Rogers Ramanujan type sums.
Here we propose to check the Hatayama et al. conjecture, using Lie algebraic theoretic methods.
We use 
Freudenthal's formula, which we computerized, to verify the identities for all the 
algebras at low rank and low level. We find complete agreement with the conjecture.
\endpage 

One of the most intriguing and interesting mathematical identities are the ones found by
Ramanujan and proved by Rogers and their generalizations, known as Generalized Rogers
Ramanujan identities (GRR). The GRR are a cornerstone in number theory, partitions theories
and combinatorics.

Amazingly, it turns out,  that the GRR identities play a pivotal role in the physics of two
dimensional systems, and can be explained, and proved, by considering central physical
systems. The GRR arose in physics first in two apparently disconnected systems.
In 1980 Baxter 
\REF\Baxter{R.J. Baxter, Exactly solved models is statistical mechanics (Dover books on
physics).}\r\Baxter\ 
observed that in the calculation of the local state probabilities in the Hard
Hexagon model, the result gives precisely the Ramanujan expressions,
which was essential in Baxter's calculations. In a separate development Lepowsky and Primc
\REF\Lep{J. Lepowsky and M. Primc, Contemporary Mathematics vol. 46 (AMS, Providence,1985).}
\r\Lep\
expressed the string functions  of $SU(2)$ affine algebras as GRR type
sums. These were later shown to be the characters of parafermionic field theories
\REF\GepQiu{D. Gepner and Z. Qiu, Nucl. Phys. B 285 (1987) 423.}\r\GepQiu.
 In fact, these two appearances of GRR, are related as it was
shown that the local state probabilities are given by the characters of the fixed point conformal
field theories \REF\MelGep{E. Melzer, Int. J. Mod. Phys. A 9 (1994) 1115; D. Gepner, Phys. Lett B 348 (1995) 377.}\r\MelGep.
 Thus, the conjecture that the characters of
any conformal field theory, which is the fixed point of an RSOS lattice model, can be
expressed as GRR type sums.
Several examples of  characters in CFT were studied, and were  shown to be
expressible as Rogers Ramanujan type sums 
\REF\Das{V. S. Dasmahapatra, T. R. Klassen, B. M. McCoy and E. Melzer,
J. Mod. Phys. B7 (1993) 3617}
\REF\Kedem{R. Kedem, T. R. Klassen, B. M. McCoy and E. Melzer, Phys. Lett. B 307
(1993) 68.}
\REF\Baver{E. Baver and D. Gepner, Phys. Lett. B 372 (1996) 231.}
 \REF\Kuniba{A. Kuniba, T. Nakanishi and J. Suzuki, Mod. Phys. Lett A8 (1993)1649.}
 \REF\BG{A. Belavin
and D. Gepner, Lett. Math. Phys. 103 (2013) 1399.}
\REF\GG{A. Genish and D. Gepner, Nucl. Phys. B, 886 (2014) 554.}
\r{\Das-\GG}. 

\def\bfn{{\bf n}}

Rogers Ramanujan type sums are of the form,
$$\sum_{\bf n}  {q^{\bfn  B  \bfn+A\bfn} \over (q)_{\bf n}},\e$$
where $B$ is some matrix, $A$ is a vector and $\bf n$ is a vector of non negative integers.
We used the Pochhammer symbol defined as
$$(q)_{\bf n}=\prod_{i=1}^m (q^{r_i})_{n_i},\qquad (q)_n=\prod_{j=0}^n (1-q^j).\e$$
There might be some restriction on the sum and $r_i$ are some integers. 

The simplest cases were found by Slater 
\REF\Slater{L.J. Slater, Proc. Lond. Math. Soc. 54 (1953) 147.}\r\Slater,
and it gives  GRR expressions for the characters of the Ising model. For example,
$$\chi_0(q)+\chi_{1\over2}(q)=\prod_{j=0}^\infty (1+q^{j+1/2})=\sum_{n=0}^\infty {q^{n^2/2}\over (q)_n}.\e$$

We wish to generalize the Slater identities to parafermionic field theories which are generalizations
of the Ising model. In the work 
\REF\Gep{D. Gepner, Lett. Math. Phys. 105 (2015) no. 6, 769.}\r\Gep\
 we presented GRR identities which express the
characters of generalized parafermions 
\REF\GepPF{D. Gepner, Nucl. Phys. B 290 (1987) 10.}\r\GepPF.
The characters of the parafermions are
the string functions of untwisted affine algebras. Here, we describe a generalization of these
identities to the string functions of twisted affine algebras. These are the characters
of oribifolds of parafermionic field theories.

Hatayama et al. proposed a generalized Rogers Ramanujan conjecture for the unit fields of
twisted affine algebras  
\REF\Kun{G. Hatayama, A. Kuniba, M. Okado, T. Takagi and Z. Tsuboi,
Prog. Math. Phys. 23 (2002) 205.}\r\Kun.
Our  aim is to verify this conjecture. The twisted affine algebras were classified by Kac
\REF\Kac{V.G. Kac, Infinite dimensional Lie algebras, Cambridge Univ. press, 1990.}\r\Kac. 
 They are, $G^{(r)}_m$ where $r=2$ or $3$ is the twist type and $m$ is the rank of the algebra. 
The algebras are
$A^{(2)}_2$,
$A^{(2)}_{2 l}$ (for $l\geq2$), $A^{(2)}_{2l-1}$ (for $l\geq3$), $D^{(2)}_{l+1}$ (for $l\geq2$), $E_6^{(2)}$ and $D_4^{(3)}$. The algebra that
remains invariant under the twist is labeled by $G_{\bar 0}$ and it is the same
as the algebra obtained by removing the $\alpha_0$ node, labeled by $\bar G$, except for 
$A^{(2)}_{2 l}$, where $G_{\bar 0}$ is $B_l$ and $\bar G$ is $C_l$. The algebra $G_{\bar 0}$ is,

$A^{(2)}_2:\  A_1$,\quad
$A^{(2)}_{2l}:\ B_l$,\quad
$A^{(2)}_{2l-1} :\  C_l$,\quad
$D^{(2)}_{l+1} : \ B_l$,\quad
$E_6^{(2)}:\  F_4$,\quad
$D_4^{(3)} :\  G_2$,
where the first entry is the affine algebra and the second is the finite algebra $G_{\bar 0}$.

We label by $\alpha_a$, $a=1,2,\ldots,l$, the simple roots of $G_{\bar 0}$. The scalar product $\alpha_a\cdot\alpha_b$
of the roots is normalized such that the long roots have a square which is $2r$. We define the 
array 
$$K_{ij}^{ab}=\half \alpha_a\cdot \alpha_b \left[\min(i,j)-i j/m\right] \e $$
where $m$ is the level of the representations and $i,j=1,2,\ldots ,m-1$.
We use also $a_0$ (the mark of the zero root) which is always equal to one, except for $A^{(2)}_{2l}$, where it is 
equal to $2$. 
We define $t_a={\rm max}(\alpha_a^2/2,a_0)$, for each of the roots of $G_{\bar 0}$.
We denote by $M$ the root lattice of $G_{\bar 0}$ and define the map
$${\cal I} (\alpha_a) =\epsilon_a \alpha_a,\e $$
where $\epsilon_a$ is equal to $1$, except for $A^{(2)}_{2 l}$ and $a=l$ (the long root), where it is $2$.

We also use the Pochhammer symbol 
$$(q)_n=\prod_{j=1}^n (1-q^j)$$
and define $q_a=q^{t_a}$. We define the set of non--negative integers
$$n^{(a)}_j,\e$$
where $a=1,2,\ldots,l$ and $j=1,2,\ldots,m-1$. We set
$$(q)_{\bf n}=\prod_{a,j} (q_a)_{n^{(a)}_j}\e$$

We are ready now to state the conjecture of Hatayama  et al. \r\Kun.
We define the GRR sum
$$N_{m,\lambda}(q)=\prod_{a=1}^ m (q_a)_{\infty}^{-1} \sum_{n^{(a)}_j\geq 0} {q^{{\cal L}({\bf n})}\over (q)_{\bf n}},\e$$
where 
$${\cal L}({\bf n})=\sum_{a,b,i,j} n^{(a)}_i n^{(b)}_j K^{ab}_{ij}.\e$$
The sum in eq. (8) is limited to $n_j^{(a)}$ which obey,
$$\sum_{a=1}^l \sum_{j=1}^{m-1} j \alpha_a n^{(a)}_j={\cal I}(\lambda) \mod m M. \e$$

The string function $c^0_\lambda(q)$ is then given by
$$c^0_\lambda(q)=N_{m,\lambda}(q^{1/{a_0}}),\e$$
for any weight $\lambda$.
This equation (11) is the Rogers Ramanujan type identity for the string functions, as conjectured by Hatayama et al..
$c^0_\lambda(q)$ is the string function for the identity representation, $\Lambda=0$ and the weight 
$\lambda$. The string functions are defined by
$$c^\Lambda_\lambda(q)=q^\kappa \sum_{n=0}^\infty \,{\dim}\, V_{\lambda+n\delta} \, q^n ,\e$$
in the representation with highest weight $\Lambda$. Here $\delta$ is measuring the grade,
and $\kappa$ is some number. 

Our aim in this paper is to check systematically the GRR expressions above, eq. (11). Since the string 
functions are the generating functions for the multiplicities of weights in the affine
representations, we can use algebraic means to calculate them, grade by grade.
We find, of particular suitability, the formula by Freudenthal \r\Kac\ which is
$$\left( |\Lambda+\rho|^2-|\lambda+\rho|^2\right){\rm dim} V_\lambda=2 \sum_{\alpha\in\Delta_+}\sum_{j\geq1} ({\rm mult\,}
\alpha) (\lambda+j\alpha|\alpha)\, {\rm dim} V_{\lambda+j\alpha},\e$$
where $\Lambda$ is the highest weight, $\lambda$ is the weight, $\rho$ is half the sum of 
positive roots, $\Delta_+$ is the set of positive roots and ${\rm dim} V_\lambda$ is the
multiplicity of the weight $\lambda$. In  Freudenthal's  formula, we use the twisted affine structures.
Then Freudenthal's formula can be used recursively, grade by grade, to calculate all the multiplicities
up to some grade. We implemented this algorithm in the computer program ALGEBRA .
This  GRR conjecture applies only to the singlet representation, $\Lambda=0$.
Using the program ALGEBRA we checked this conjecture for many algebras, in particular all
the rank two, three and four algebras, at the levels $m=1,2,3,4$. Quite amazingly this conjecture
gives indeed the correct string functions.

Let us give now some examples. 
At level one the string functions are known rigorously
\REF\KacPet{V. Kac and D. Petersen, Bull.A.M.S. 3 (1980) 3.}\r\KacPet,
and they agree with the conjecture of 
Hatayama et al.
At level two, the expressions for the GRR sums assume
a particular simple form. The expression can be written as,
$$\chi^0_\lambda (q^{a_0})\prod_i (q^{t_i})_\infty =\sum_{m_i\geq 0 \atop m_i=\epsilon_i\mod 2} {q^{mCm/4}\over \prod_i (q^{t_i})_{m_i}},\e$$
where $C$ is the matrix of scalar products of $G_{\bar 0}$, $C=\alpha_i\cdot \alpha_j$ and $t_i={\rm max}(\alpha_i^2/2,a_0)$. Here $\epsilon_i=0,1$ and 
$${\cal I}(\lambda)=\sum_i \epsilon_i\alpha_i.\e$$
$\chi^0_\lambda$ is the string function for the representation with highest weight $0$ and weight $\lambda$.

Let us take the case of $A^{(2)}_2$ at level $2$. Here the matrix $C$ is
$$C=\pmatrix{4 & -2\cr -2 & 2 \cr},\e$$
and $t_1=t_2=2$, and $a_0=2$. So we have, by calculating eq. (14), with $\epsilon_i=0$,
$$\chi^0_0(q^2)=1 + 4 q^2 + 14 q^4 + 40 q^6 + 104 q^8 + 248 q^{10} + 556 q^{12} + 
 1184 q^{14} + 2421 q^{16} +\ldots\e$$
Using the program ALGEBRA we get precisely this series for the string function up to order $8$.

Let us give another example at rank two, which is $D_3^{(2)}$. Here the matrix $C$ is again as
in eq. (16) but $t_1=2$ and $t_2=1$ and $a_0=1$. From eq. (14) we find
$$\chi_0^0(q)=1 + q + 5 q^2 + 8 q^3 + 24 q^4 + 39 q^5 + 90 q^6 + 147 q^7 + 
 297 q^8 + 477 q^9 +\ldots\e$$
 where we take $\epsilon_i=0$.
 From ALGEBRA we get exactly this expansion up to order $9$.
 Taking $\epsilon_2=1$ and $\epsilon_1=0$ we find the string function
$$\chi^0_{\alpha_2}(q)=1+2 q +7 q^2+13 q^3+32 q^4+57 q^5+119 q^6+204 q^7+385 q^8 \ldots\e$$
The calculation of the string function from ALGEBRA agrees with this.

 Another rank two example is $D_4^{(3)}$. Here the scalar product matrix is
 $$C=\pmatrix{6 & -3 \cr -3 & 2\cr},\e$$
 and $t_1=3$, $t_2=1$ and $a_0=1$. We find from eq. (14) that,
$$\chi^0_0(q)=1 + q + 5 q^2 + 10 q^3 + 21 q^4 + 42 q^5 + 83 q^6 + 143 q^7 + 
 263 q^8 + 448 q^9+\ldots\e$$ 
 where we take $\epsilon_i=0$.
Again, from ALGEBRA we get exactly this string function up to grade $9$.
Taking $\epsilon_2=1$ and $\epsilon_1=0$ we find from our GRR,
$$\chi_{\alpha_2}^0(q)=1+3 q+6 q^2+15 q^3+31 q^4+57 q^5+110 q^6+198 q^7+338 q^8+\ldots\e$$
which is also what we obtain from ALGEBRA. 

Let us move now to rank three. Our first example is $A_5^{(2)}$. Here the matrix $C$ is
$$C=\pmatrix{2 & -1& 0\cr
                       -1&2 & -2\cr
                       0&-2 &4\cr},\e$$
and $t_1=t_2=1$ and $t_3=2$.  From eq. (14) we find, taking $\epsilon_i=0$,
$$\chi_0^0(q)=    1 + 2 q + 12 q^2 + 33 q^3 + 108 q^4 + 269 q^5 + 699 q^6 + 1593 q^7 + 
 3640 q^8 + 7717 q^9+\ldots\e$$        
 Amazingly, from ALGEBRA we get exactly this string function.    
 We find for the string function,
 $\chi_{\alpha_1}^0(q)$, where $\lambda=\alpha_1$, the first simple root, from our GRR expression, eq. (14),
 $$\chi_\lambda^0(q)=1+5 q+19 q^2+60 q^3+169 q^4+436 q^5+1055 q^6+2419 q^7+5309 q^8+\ldots\e$$
 Here we take $\epsilon_1=1$ and $\epsilon_2=\epsilon_3=0$.
 From ALGEBRA we find exactly this string function up to this order.
 
 Let us consider now the algebra $E_6^{(2)}$, which is at rank four, and  at the level two. The matrix of
 scalar products is
 $$C=\pmatrix{ 4 & -2  &0&0\cr
                         -2&4&-2&0\cr
                         0&-2&2&-1\cr
                         0&0&-1&2\cr}.\e$$
Thus $t_1=t_2=2$ and $t_3=t_4=1$. Taking $\epsilon_i=0$ we find from the GRR,
$$\chi_0^0(q)=1 + 2 q + 19 q^2 + 58 q^3 + 234 q^4 + 644 q^5 + 1944 q^6 + 4874 q^7 + 
 12559 q^8+\ldots\e$$
 which is identical to the calculated affine string function. By taking in the GRR, $\epsilon_4=1$
 and the rest of $\epsilon_i$ zero, we get
 $$\chi_{\alpha_4}^0(q)=1+7 q+32 q^2+119 q^3+386 q^4+1133 q^5+3081 q^6+7884 q^7+19171 q^8+\ldots\e$$
 which is again confirmed by ALGEBRA.
 
 Let us turn now to examples with levels greater than $2$.  Here we use the GRR conjecture
 eqs. (8,11). Consider the algebra $D_4^{(3)}$ at level three. The matrix of scalar product is given
 in eq. (20).  By a direct calculation we find for the string function,
  $$\chi_0^0(q)=1 + q + 5 q^2 + 15 q^3 + 33 q^4 + 78 q^5 + 179 q^6 + 350 q^7 + 
 706 q^8 + 1366 q^9+\ldots\e$$
 Using ALGEBRA we find the same string function. At level $4$ for the same algebra we find,
 $$\chi_0^0(q)=1 + q + 5 q^2 + 15 q^3 + 41 q^4 + 100 q^5 + 251 q^6 + 546 q^7 + 
 1183 q^8 + 2453 q^9+\ldots\e$$
 Again, this equation is confirmed by the computer program we wrote.

This work  is centered around the Generalized Rogers Ramanujan identities for
the twisted affine algebras. First, we wish to find a generalization of these identities to the
non singlet representations. In ref. \r\Gep\ we found many such generalizations for the
untwisted algebras. Unfortunately, the naive guess does not work for the twisted algebras, but we
are confident that with more work, such GRR expressions could be found.

Our second objective is to prove these GRR identities. We already have encouraging results 
for the level two cases. We believe that the $q$--diagram technique, that we used to prove the level
two untwisted cases
\REF\Genish{A. Genish and D. Gepner, Nucl.Phys. B 897 (2015) 179; Nucl. Phys. B 907 (2016) 154.}
\r\Genish,
can be adapted to work also for the twisted case, and be used to prove and generalize the 
Rogers Ramanujan type identities.

\ack
We thank A. Abouelsaood for his colaboration on the computer program ALGEBRA and 
I. Deichaite for remarks on the manuscript.

\refout

\bye